\documentclass[hyper]{JHEP3}
\pdfoutput=1
\usepackage{amsmath,amssymb,epsfig}
\usepackage{amsfonts}
\usepackage{verbatim}
\usepackage{latexsym}
\usepackage{amsthm}
\usepackage{amsbsy}
\usepackage{multirow}

\def\one{{\hbox{ 1\kern-.8mm l}}}
\newcommand{\Dslash}{\not{\hbox{\kern-4pt $D$}}}
\newcommand{\pdslash}{\not{\hbox{\kern-2pt $\partial$}}}
\newcommand{\cL}{\mathcal{L}}

\newcommand{\Tr}{\mathrm{Tr}}

\newcommand{\SO}{\mathrm{SO}} 
\newcommand{\SU}{\mathrm{SU}} \newcommand{\U}{\mathrm{U}}

 \newcommand{\ie}{\emph{i.e.}\:}
 \newcommand{\pd}{\partial}

\newcommand{\Comment}[1]{{}}

\def\IZ{{\mathbb Z}}

\def\IR{{\mathbb R}}

\newcommand{\bc}{\begin{center}}
\newcommand{\ec}{\end{center}}
\newcommand{\ba}{\begin{array}}
\newcommand{\ea}{\end{array}}
\newcommand{\beq}{\begin{equation}}
\newcommand{\eeq}{\end{equation}}
\newcommand{\bea}{\begin{eqnarray}}
\newcommand{\eea}{\end{eqnarray}}
\newcommand{\bmx}{\begin{pmatrix}}
\newcommand{\emx}{\end{pmatrix}}
\newcommand{\nn}{\nonumber}

\newcommand{\be}{\begin{equation}}
\newcommand{\ee}{\end{equation}}

\newcommand{\m}{\mu}

\newcommand{\del}{\partial}
\newcommand{\half}{\frac{1}{2}}

\newcommand{\tphi}{{\tilde\phi}}

\newcommand{\eref}[1]{Eq.~(\ref{#1})}

\newcommand{\hD}{{\hat{D}}}

\newcommand{\cG}{{\cal G}}
\newcommand{\tcG}{{\tilde{\cal G}}}

\def\IB{\relax{\rm I\kern-.18em B}}
\def\IC{{\relax\hbox{\kern.3em{\cmss I}$\kern-.4em{\rm C}$}}}
\def\ID{\relax{\rm I\kern-.18em D}}
\def\IE{\relax{\rm I\kern-.18em E}}
\def\IF{\relax{\rm I\kern-.18em F}}
\def\II{\relax{\rm I\kern-.18em I}}
\def\IZ{\relax{\sf Z\kern-.35em Z}}
\def\Id{\relax{1\kern-.32em 1}}
\def\IG{\relax\hbox{$\inbar\kern-.3em{\rm G}$}}
\def\IR{\relax{\rm I\kern-.18em R}}

\normalsize

\title{D2 to D2} 

\author{Bobby Ezhuthachan\footnote{email:
    bobby@theory.tifr.res.in}\, , Sunil Mukhi\,\footnote{email:
    mukhi@tifr.res.in}\, and Constantinos Papageorgakis\,
  \footnote{email: costis@theory.tifr.res.in}\\ \it Tata Institute of
  Fundamental Research,\\ \it Homi Bhabha Rd, Mumbai 400 005, India}

\abstract{Starting from maximally supersymmetric (2+1)d Yang-Mills
theory and using a duality transformation due to de Wit, Nicolai and
Samtleben, we obtain the ghost-free Lorentzian 3-algebra theory that
has recently been proposed to describe M2-branes. Our derivation does
not invoke any properties of 3-algebras. Being derivable from SYM, the
final theory is manifestly equivalent to it on-shell and should not be
thought of as the IR limit that describes M2-branes, though it does
have enhanced R-symmetry as well as superconformal symmetry
off-shell. }

\preprint{TIFR/TH/08-22}


\begin{document}

\section{Introduction}

There has been intense recent activity regarding a certain class of
$\mathcal N=8$ superconformal theories in three dimensions, following
the work of Bagger-Lambert (BL)
\cite{Bagger:2006sk,Bagger:2007jr,Bagger:2007vi} and also Gustavsson
\cite{Gustavsson:2007vu}, as these theories are potential candidates
for the worldvolume description of multiple M2-branes in
M-theory. These constructions rely on the introduction of an algebraic
structure going under the name of a Lie 3-algebra, which is necessary
for the closure of the supersymmetry algebra. The metric versions of
the above theories\footnote{See also \cite{Gran:2008vi} for the
treatment of a non-metric proposal.} fall into two classes, depending
on whether the invariant bilinear form in 3-algebra space is positive
definite or indefinite: the Euclidean theories originally proposed by
Bagger-Lambert and their more recent Lorentzian counterparts
\cite{Gomis:2008uv,Benvenuti:2008bt,Ho:2008ei}. 

The Lorentzian 3-algebra theories have been claimed to be capturing
the low-energy worldvolume dynamics of multiple parallel M2-branes but
are plagued by apparent unitarity problems due the presence of
ghost-like degrees of freedom in the classical action. In order to
address this issue, a proposal has appeared which enlarges the theory
by gauging a shift symmetry for one of two ghosts, via the
introduction of appropriate gauge fields. This construction leads to a
manifestly ghost-free
spectrum~\cite{Bandres:2008kj,Gomis:2008be}. However, this results in
the other ghost field being frozen to a constant vev. Then, as already
observed in \cite{Ho:2008ei} along the lines of \cite{Mukhi:2008ux},
the theory reduces precisely to maximally supersymmetric Yang-Mills in
three dimensions with a gauge coupling equal to the scalar
vev.\footnote{A similar procedure has been carried out in the context 
of Janus field theory in Ref.~\cite{Honma:2008un}.} This
and other properties have been used to argue \cite{Gomis:2008be} that
the ghost-free Lorentzian 3-algebra theory is indeed the IR limit of
SYM. 

However,\footnote{This point was stressed in
Ref.~\cite{Bandres:2008kj}.} such a precise reduction should make one
suspicious that the ghost-free Lorentzian 3-algebra is the {\it same}
theory as SYM rather than its infrared limit. In this letter we would
like to reinforce this interpretation by reversing the procedure of
\cite{Ho:2008ei}. We will show that starting from $\mathcal N=8$ SYM one
can systematically - and uniquely - recover the theory of
\cite{Bandres:2008kj,Gomis:2008be}. 

Let us summarise how we will achieve the transformation of SYM theory
into the ghost-free Lorentzian 3-algebra theory. First we will use a
prescription for dualising non-abelian gauge fields in the special
case of three dimensions, due to de Wit, Nicolai and Samtleben (dNS)
\cite{Nicolai:2003bp,deWit:2003ja,deWit:2004yr}. In this prescription the
gauge field $A_\mu$ gets replaced by two non-dynamical gauge fields
$A_\mu,B_\mu$ with a $B\wedge F(A)$ type kinetic term, plus an extra
scalar which ends up carrying the dynamical degree of freedom of the
original YM gauge field. Once this is done, we observe a potential
$\SO(8)$ symmetry in the theory under which the extra scalar mixes
with the seven existing ones. We realise this $\SO(8)$ symmetry as a
formal symmetry (acting also on coupling constants) by replacing
$g_{YM}$ with an $\SO(8)$ vector of coupling
constants $g_{YM}^I$. Finally, the latter is promoted to a scalar
field that is an $\SO(8)$ vector, whose equations of motion render it
constant. We justify all these steps and note that they do not change
the on-shell theory in any way. However, off-shell they give a theory
with enhanced symmetries: $\SO(8)$ R-symmetry instead of $\SO(7)$ and
superconformal symmetry instead of ordinary supersymmetry. We also
comment on the construction of $\SO(8)$-covariant, gauge-invariant
operators and find that, unsurprisingly, these are only present
off-shell and reduce to an $\SO(7)$-covariant basis on any physical
solution.

Even though the above procedure closely follows the treatment of
\cite{Gomis:2008be}, albeit in the opposite order, we believe that
this angle will help to demystify the connection between the two
theories and clarify that the resultant Lagrangean is nothing but a
re-writing of maximally supersymmetric Yang-Mills
theory. Interestingly, this re-writing allows one to recover the
conformal and $\SO(8)$ symmetries  off shell, which are however
spontaneously broken by any physical vev of the theory. The authors
of \cite{Gomis:2008be} propose a prescription for recovering the
$\SO(8)$ R-symmetry by integrating over all values of the constrained
ghost field. In our interpretation this amounts to integrating SYM
theory over all values of the coupling constant. This seems unnatural
at best and should also lead to violations of basic QFT axioms, such
as locality, and result into problems with cluster decomposition. In
that sense, starting from the theory of D2-branes, one ends up with the
theory of D2-branes. The M2's only emerge in the limit of taking the
SYM coupling to infinity, as is of course well-known.

Our discussion has no bearing on the original ``un-gauged'' Lorentzian
BL proposal of \cite{Gomis:2008uv,Benvenuti:2008bt,Ho:2008ei}, which
could still be a non-trivial example of an $\SO(8)$-invariant
theory. However, since this theory still needs to be demonstrated to
be free of ghosts, we believe that promising candidates for the
worldvolume theory of multiple M2-branes in noncompact space must lie
elsewhere.

The rest of this note is organised as follows. In the next section we
present our main argument in full detail. We proceed with a discussion
on the $\SO(8)$-covariant gauge-invariant operators in section 3. In
section 4 we propose a potential generalisation of the
dNS duality to four dimensions and speculate that it
might be useful in studying 4d dualities. We close in section 5 with a
discussion of our results.

\section{From SYM to Lorentzian 3-algebras}

We start with the maximally supersymmetric interacting super
Yang-Mills Lagrangean in 2+1 dimensions based on an arbitrary Lie
algebra ${\cal G}$:
\begin{equation}\label{sym}
\begin{split}
  \cL = &\Tr\left( -\frac{1}{4g^{2}_{YM}}
F_{\mu\nu}F^{\mu\nu} - \frac{1}{2}D_\mu
    X^iD^\mu X^i - \frac{g_{YM}^2}{4}[X^i,X^j][X^j,X^i]\right.\\
  &\left.\qquad \qquad\qquad+\frac{i}{2}\bar \Psi \Dslash\Psi +
    \frac{i}{2}g_{YM}\bar \Psi \Gamma_i[X^i,\Psi]\right)\;,
\end{split}
\end{equation}
Here $A_\mu$ is a gauge connection on ${\cal G}$. The field
strength and the covariant derivatives are defined as:
\begin{equation}
\label{fscd}
  F_{\mu\nu} = \partial_\mu A_\nu - \partial_\nu A_\mu - 
  [A_\mu,A_\nu]\qquad \textrm{and}\qquad D_\mu = \partial_\mu -
  [A_\mu,\cdot\;]\;.
\end{equation} 
The $X^i$s are seven matrix valued scalar fields transforming as
vectors under the $\SO(7)$ R-symmetry group. The $\Psi$s are
two-component spinors in (2+1)d and also 8-component spinors of
$\SO(7)$.

When ${\cal G}$ is $\U(N)$ this theory is the low energy worldvolume
action for multiple parallel D2-branes in flat space. For the other
classical Lie algebras, it describes D-branes at orientifolds. Our
goal in this note is to show that for any gauge group, this Lagrangean
can be brought to the form of the Lorentzian Bagger-Lambert or
3-algebra field theory proposed in
\cite{Gomis:2008uv,Benvenuti:2008bt,Ho:2008ei}, or more precisely to
the ``gauged'' version of the above theory described in
\cite{Bandres:2008kj,Gomis:2008be}. 

We proceed by introducing two new fields $B_{\mu}$ and $\phi$ that are
adjoints of ${\cal G}$. In terms of these new fields the
dNS duality transformation
\cite{Nicolai:2003bp,deWit:2003ja,deWit:2004yr} is the replacement:
\begin{equation}\label{dual}
\Tr\left( -\frac{1}{4 g_{YM}^2}
  F^{\mu\nu}F_{\mu\nu}\right) \rightarrow \Tr \left(
  \frac{1}{2}\epsilon^{\mu\nu\lambda}B_\mu F_{\nu\lambda}
  -\frac{1}{2}\left(D_{\mu}\phi- g_{YM} B_{\mu} \right)^2\right)\;.
\end{equation}
We see that in addition to the gauge symmetry $\cG$, the new action
has a {\it noncompact abelian} gauge symmetry that we can call $\tcG$,
which has the same dimension as the original gauge group
$\cG$.\footnote{For the $\U(1)$ case the scalar $\phi$ is a periodic
field. For non-simply connected gauge groups, \ie for $\U(N)$, the
$\U(1)$ part will still have the aforementioned periodic shift
symmetry. Our discussion here applies to the $\SU(N)$ part of the
theory after decoupling the $\U(1)$ supermultiplet, so compactness is
not an issue.} This symmetry consists of the transformations:
\be
\label{ncgt}
 \delta \phi = g_{YM}M \;,\qquad \delta B_\mu = D_\mu M\;,
\ee
where $M(x)$ is an arbitrary matrix, valued in the adjoint of ${\cal
G}$. Clearly $B_{\mu}$ is the gauge field for the shift symmetries
$\tilde{\cal G}$. Note that both in \eref{dual} and \eref{ncgt}, the
covariant derivative $D_\mu$ is the one defined in \eref{fscd}.

If one chooses the gauge $D^\mu B_\mu=0$ to fix the shift symmetry,
the degree of freedom of the original Yang-Mills gauge field $A_\mu$
can be considered to reside in the scalar $\phi$. In this sense one
can think of $\phi$ as morally the dual of the original
$A_\mu$~\cite{Nicolai:2003bp,deWit:2003ja,deWit:2004yr}.
Alternatively we can choose the gauge $\phi=0$, in which case the same
degree of freedom resides in $B_\mu$.  The equivalence of the RHS to
the LHS of \eref{dual} can be conveniently seen by going to the latter
gauge. Once $\phi=0$ then $B_\mu$ is just an auxiliary field and one
can integrate it out to find the usual YM kinetic term for
$F_{\mu\nu}$.\footnote{If the scalar field $\phi$ were not introduced,
the duality would go through but there would be no gauge symmetry
associated to the vector field $B_\mu$, and it would simply end up as
an auxiliary field in an uninteresting field theory. Here on the other
hand, we instead get an interesting dual field theory containing all
of $A,B,\phi$.}

We can now proceed to study the dNS-duality transformed $\mathcal N
=8$ Yang-Mills theory. Its Lagrangean is:
\begin{equation}\label{D2}
\begin{split}
  \cL = &\Tr\left(\frac{1}{2}
      \epsilon^{\mu\nu\lambda}B_{\mu}F_{\nu\lambda} -
    \frac{1}{2} \left(D_{\mu}\phi -g_{YM}B_{\mu}\right)^{2}
    -\frac{1}{2}D_\mu X^i D^\mu X^i \right.  \\ 
\qquad &\left.-
    \frac{g_{YM}^2}{4}[X^i,X^j][X^j,X^i] +\frac{i}{2}\bar \Psi
    \Dslash\Psi + \frac{i}{2}g_{YM}\bar \Psi
    \Gamma_i[X^i,\Psi]\right)\;.
\end{split}
\end{equation}

The gauge-invariant kinetic terms for the eight scalar fields have a
potential $\SO(8)$ invariance, which can be exhibited as follows. First
rename $\phi(x) \to X^8(x)$. Then the scalar kinetic terms become
$-\half \hD_\mu X^I \hD^\mu X^I$, where:
\bea
\hD_\mu X^i &=& D_\mu X^i = \del_\mu X^i -  [A_\mu,X^i],
\quad i=1,2,\ldots,7\nn\\
\hD_\mu X^8 &=& D_\mu X^8 - g_{YM}B_\mu = \del_\mu X^8 -  [A_\mu,X^8]
- g_{YM} B_\mu\;.
\eea
Defining the constant 8-vector:
\beq
\label{gvec}
g_{YM}^I = (0,\ldots,0,g_{YM})\;,\quad I=1,2,\ldots,8\;,
\eeq 
the covariant derivatives can together be written:
\beq
\hD_\mu X^I = D_\mu X^I - g_{YM}^I B_\mu\;.
\eeq

One can now uniquely write the SYM action in a form that is
$\SO(8)$-invariant under transformations that rotate both the fields
$X^I$ and the coupling-constant vector $g_{YM}^I$:
\begin{equation}\label{so8} 
\begin{split}
  \cL =&  \Tr\Big( \frac{1}{2}\epsilon^{\mu\nu\lambda}B_\mu F_{\nu\lambda}
   - \frac{1}{2} \left( D_\mu X^I -g_{YM}^I B_\mu
  \right)^2 +\frac{i}{2}\bar \Psi \Dslash\Psi +
    \frac{i}{2}g_{YM}^I\bar \Psi \Gamma_{IJ}[X^J,\Psi]\\
&- \frac{1}{12} \left( g_{YM}^I [X^J,X^K] +
    g_{YM}^J[X^K,X^I]  + g_{YM}^K[X^I,X^J]\right)^2 \Big)\;.
  \end{split}
\end{equation}
Here, in writing down the Yukawa-type interaction we
have embedded the $\SO(7)$ $\Gamma$-matrices into $\SO(8)$ using
$\Gamma^i = \tilde\Gamma^8 \tilde \Gamma^i$. On the RHS the
$\Gamma$-matrices are those of $\SO(8)$. One can easily see that with
this definition the LHS matrices satisfy the Clifford algebra of
$\SO(7)$.

The $\mathcal N=8$ supersymmetry transformations for the theory above can also
be written in a formally $\SO(8)$-invariant form:
\begin{eqnarray}
  \delta X^I=i\bar\epsilon\;\Gamma^{I}\Psi \;, & \qquad & \delta\Psi =
  \left(D_{\mu}X^I - g_{YM}^I B_{\mu} 
  \right)\Gamma^{\mu}\Gamma_{I}\epsilon -\frac{1}{2}
  g_{YM}^I [X^J,X^K]\,\Gamma_{IJK}\epsilon  \nonumber \\
  \delta A_{\mu}=
  \frac{i}{2}g_{YM}^I\bar{\epsilon}\;\Gamma_{\mu}\Gamma_{I}\Psi\;,
  &\qquad &
  \delta B_{\mu} =  i\bar{\epsilon}\;\Gamma_{\mu}\Gamma_{I}[X^I,\Psi]  \;.
\end{eqnarray}

The theory in \eref{so8} is merely a re-writing of the dNS-transformed
$\mathcal N=8$ SYM action. It has the nice property that $g_{YM}^I$
can be an arbitrary 8-vector, not necessarily of the form in
\eref{gvec}, with the theory depending only on its norm.  To see this,
simply pick an arbitrary vector $g_{YM}^I$ with $\sqrt{g_{YM}^I
  g_{YM}^I}=g_{YM}$ and use the $\SO(8)$ invariance to rotate to a
basis where it takes the form \eref{gvec}. It is in this
basis that the field $\phi$ appearing in the dNS duality
transformation can be identified with $X^8$.

At this stage the theory is only formally $\SO(8)$ invariant, as the
transformations must act on the coupling constants as well as the
fields. So $\SO(8)$ is not a true field-theoretic
symmetry.\footnote{An equivalent way to say this is that any
particular choice of the coupling constant vector breaks $\SO(8)$ down
to the $\SO(7)$ orthogonal to it.} By the same token, the theory is
formally conformal invariant if, in addition to scaling the fields,
one scales the dimensional coupling constant vector $g_{YM}^I$ (this
fact is inherited from the original SYM in (2+1)d which already had
this property). Again the conformal invariance is not a true
field-theoretic symmetry -- this is particularly obvious, as $\mathcal
N=8$ SYM at finite coupling can hardly be conformal!

However at this stage we are in a position to simultaneously convert
the formal $\SO(8)$ and conformal symmetries to true field-theoretic
symmetries by replacing the coupling constant vector by a set of eight
new scalar fields $X_+^I(x)$. The resulting theory will be on-shell
equivalent to the theory in \eref{so8} if and only if the new scalar
field $X_+^I(x)$ has an equation of motion that renders it
constant. Then we can simply write an arbitrary classical solution as
$\langle X_+^I\rangle = g_{YM}^I$ and the theory will reduce to that
in \eref{so8}, which we have already shown to be correct.

To enforce the constancy of $X^{I}_{+}$ one introduces an
auxiliary Lagrange multiplier. This can be implemented by adding
the following term to the action:
\be
\cL_{\rm{mult}}  = C^{\mu}_{I}\pd_{\mu}X^{I}_{+} \;.
\ee
As pointed out in footnote 5, it is typically not useful to introduce
a vector field without an associated gauge symmetry. To see what goes
wrong for the case at hand, note that the above term has a new
symmetry~\cite{Gomis:2008be}:
\be\label{gauge1}
\delta C^\mu_I = \epsilon^{\mu\nu\lambda}\pd_\nu \alpha_{\lambda\;I}
\ee
that needs to be gauge fixed. On doing this, one finds (for more
details, see Ref.~\cite{Gomis:2008be}) a
standard gauge kinetic term:
\be
\cL_{\rm{GF}} = (\epsilon_{\mu\nu\lambda} \pd^\nu C^{\lambda I})^2 =
(\pd^\nu C^{\lambda I}-\pd^\lambda C^{\nu I})^2 = (F_{\nu\lambda}(C^I))^2\;,
\ee
where $F_{\nu\lambda}(C^I)$ denotes the field strength of $C^I_\mu$.
Having obtained a kinetic term, the $C^I_\mu$'s will introduce new
negative norm states through their $C^I_0$ component. 

Therefore we impose a shift symmetry for $C_\mu^I$ by introducing a
new scalar field, which we call $X^I_-$, and writing the relevant term
in the action as:
\be
\cL_{\rm{mult}}  = (C^{\mu \;I}- \pd^\mu X_-^I)\pd_\mu X^{I}_{+} \;.
\ee
This action now has an 8-dimensional abelian
(shift) symmetry:
\beq
\delta X_-^I = \lambda^I,\quad \delta C_\mu^I = \del_\mu \lambda^I\;,
\eeq
which will remove the negative-norms states associated to $C_\mu^I$.

The rest of the argument follows Ref.~\cite{Gomis:2008be}. One needs
to gauge-fix the shift symmetry and this is done by introducing the
appropriate ghosts.  The fermionic terms are easily obtained in a
similar fashion. We introduce a superpartner $\Psi_+$ for $X_+^I$ and
a Lagrange multiplier $\eta^\mu$ enforcing the condition
$\Psi_+=0$. Following the above chain of arguments, one also needs to
introduce a superpartner for $X_-^I$, called $\Psi_-$. The structure
of these terms in $\cL_{\rm{fermion}}$ will be uniquely dictated by
supersymmetry.\footnote{The corresponding supersymmetry
transformations can be found in Ref.~\cite{Gomis:2008be}.}

We have thus ended up with the action:
\be
\begin{split}
 \cL =&  \Tr\Big(\frac{1}{2}
\epsilon^{\mu\nu\lambda}B_\mu F_{\nu\lambda} 
- \frac{1}{2} \left( D_\mu X^I -X^I_+ B_\mu
  \right)^2\\
& - \frac{1}{12} \left( X^I_+[X^J,X^K] +
    X^J_+[X^K,X^I]  + X^K_+[X^I,X^J]\right)^2\Big) \\
&+  (C^{\mu \;I}-
\pd^\mu X_-^I) \pd_\mu X^{I}_{+} + \cL_{\rm{gauge-fixing}} +
\cL_{\rm{ghosts}}+  \cL_{\rm{fermions}}\;.
\end{split}
\ee
This is precisely the Bagger-Lambert action for an arbitrary
Lorentzian 3-algebra (based on a Lie algebra $\cG$). Because our
manipulations required the field $X_+^I$ to be constant on-shell, we
were led not to the original form discovered in
Refs.~\cite{Gomis:2008uv,Benvenuti:2008bt,Ho:2008ei} (which indeed has
not yet been shown to be ghost-free) but directly to the modified one
subsequently proposed in Refs.~\cite{Bandres:2008kj,Gomis:2008be}
containing the gauge field $C_\mu^I$, for which freedom from ghosts is
easily demonstrated.

Gauging $X_-^I$ to zero using the shift symmetry and eliminating
$C_\mu^I$, we obtain the desired constraint $\del_\mu X_+^I=0$, whose
general solution is $X_+^I= g_{YM}^I$ and the action reduces to that
in \eref{so8}.\footnote{The classical solution breaks $\SO(8)$ to $\SO(7)$ and
superconformal symmetry to ordinary supersymmetry. This breaking is
therefore {\it spontaneous}, which is an essential feature of this
approach.} Hence we have arrived at the above action by a series of
completely justified transformations starting from $\mathcal N=8$ SYM.

It is striking that all the interactions and consequent properties of
this action, which were originally derived using 3-algebras with a
Lorentzian metric, have been arrived at here without any reference to
3-algebras whatsoever. As an example, the $dim\cG$ abelian shift
symmetries that arise from the 3-algebra structure, as discussed in
some detail in Ref.~\cite{Benvenuti:2008bt}, are simply a basic
property of the dNS transformation in our approach. Likewise the
sextic interaction that arises via 3-algebra structure constants in
the original papers, appears in our work when we promote $g_{YM}$ to
an $\SO(8)$ vector of coupling constants and subsequently replace that
by an $\SO(8)$ vector of scalar fields.

\section{Gauge-invariant operators and $\SO(8)$ symmetry}

In Ref.~\cite{Gomis:2008be} it was noted that for this theory
operators of the form $\Tr(X^{I_1}\cdots X^{I_n})$ are not gauge
invariant under the abelian noncompact symmetry associated with
$B_\mu$. To redress this problem, the authors introduce a {\it
nonlocal} adjoint scalar field which we call $\tphi$, defined as:
\beq
\tphi(x)= \frac{1}{D^2}D^\mu B_\mu\;.
\eeq
They identify this with the magnetic dual to the non-abelian gauge
field $A_\mu$. They then define fields: 
\beq
Y^I = X^I- \tphi X_+^I 
\eeq
that are invariant under the shift gauge transformations and lead to
operators $\Tr(Y^{I_1}\cdots Y^{I_n})$. Under the noncompact gauge
transformations \eref{ncgt}, $\tphi\to
\tphi + M$. Therefore, one can choose a gauge in which $\tphi=0$ and
the $Y^I$s reduce to $X^I$s, so one does seem to recover $\SO(8)$
covariance in this fashion but at the cost of losing manifest gauge
invariance. 

The above discussion can be re-interpreted in terms of the dNS
duality.\footnote{We would like to acknowledge correspondence with the
authors of Ref.~\cite{Gomis:2008be}, based on which this section has
been revised.} Recall our definition of the scalar field $\phi$ in
\eref{dual}. Since this transforms under the shift symmetry as
$\phi\to \phi + g_{YM}M$, it follows that:
\beq
Z(x)\equiv \phi(x) - g_{YM}\tphi(x) = 
\left(\phi - g_{YM}\frac{1} {D^2}D^\m B_\mu\right)
\eeq
is gauge invariant. It is this field, rather than $\tphi$, that
unambiguously carries the single physical adjoint degree of freedom of
the original Yang-Mills gauge field after dNS duality. $Z$ is in
general nonlocal, and becomes a local field only when we choose the
gauge $\tphi=0$. When we apply our covariantisation procedure
(promoting $g_{YM}$ to an 8-vector and thence to the field $X_+^8$) we
find that $Z(x)$ combines with the remaining seven adjoint scalar
fields to form the 8-vector $X^I - \tphi X_+^I$, which is
precisely the set $Y^I$ defined above.

We see, as in our previous discussions, that these operators are $\SO(8)$
covariant only off-shell (when $X_+^I$ is still a field) but as soon
as $X_+^I$ develops a vev, the $\SO(8)$ is broken to $\SO(7)$.

\section{Four-dimensional duality?}

The dNS duality transformation is, as we have seen, particularly
useful in (2+1)d where it allowed us to relate $\mathcal N=8$ SYM to
the Lorentzian 3-algebra theory. However a variant of it can be
written down in $3+1$ dimensions, and we briefly describe it here in
the hope that it might enhance our understanding of four-dimensional
dualities. The transformation in 4d maps a Yang-Mills theory with
gauge field $A_\mu$ to a theory having three fields: a gauge field
$A_\mu$, a 2-form gauge field $B_{\mu\nu}$ and a second gauge field
$A_\mu'$, the latter two being in the adjoint of the original gauge
group $\cG$.

The map is as follows:
\begin{equation}
\label{fourddual}
\Tr\left( -\frac{1}{4 g_{YM}^2}
  F^{\mu\nu}F_{\mu\nu}\right) \rightarrow \Tr \left(
  \frac{1}{2}\epsilon^{\mu\nu\lambda\rho}B_{\mu\nu} F_{\lambda\rho}
  -\frac{1}{2}\left(D_{\mu}A_\nu' - D_\nu A_\mu'- 
g_{YM} B_{\mu\nu} \right)^2\right)
\end{equation}
where:
\beq
D_\mu A_\nu' = \del_\mu A_\nu' - [A_\mu,A_\nu']\;.
\eeq
In addition to the gauge symmetry $\cG$, this theory has an abelian 
``2-form'' gauge symmetry $\tcG$ that acts as:
\beq
\delta A_\mu' = g_{YM} M_\mu,\quad \delta B_{\mu\nu} = D_\mu M_\nu -
D_\nu M_\mu\;.
\eeq
The duality is demonstrated by using the abelian shift symmetry to
gauge $A'_\mu$ to 0 and then integrating out $B_{\mu\nu}$. It would
also be natural to extend this duality to make $A_\mu'$ into a gauge
field with an associated abelian symmetry. This can be done in an
obvious way by introducing a scalar field $\phi'$ and replacing
$A_\mu' \to A_\mu'-D_\mu\phi'$ everywhere in the above. The question
then is whether the formulation on the RHS of \eref{fourddual}, or the
generalisation of it with this extra scalar field, can teach us
something about supersymmetric field theories in 4d.

From the above discussion we see that the dNS transformation can have
a (3+1)d analogue. However, the same does not appear to be true in any
useful way for the rest of our procedure, namely lifting of the
coupling constant to a field and the consequent enhancement of
off-shell symmetries. It is a special feature of (2+1)d that the
Yang-Mills coupling $g_{YM}$ has the same canonical dimension as a
scalar field, namely $\half$ in appropriate units. In (3+1)d the
former has canonical dimension 0 and the latter has dimension 1, while
in (1+1)d the reverse is true. Therefore, lifting the coupling
constant to a field that is rendered constant by its equation of
motion seems to be natural only in (2+1)d. However it may still be
worth exploring whether the 4d duality transform exhibited here has
some useful application.

\section{Discussion}

The reduction of a proposed M2-brane field theory to Yang-Mills was
proposed in Ref.~\cite{Mukhi:2008ux} in the context of the
Bagger-Lambert $\mathcal A_4$ theory. There, giving a vev to a scalar
field reduces the field theory to a maximally supersymmetric
YM theory plus corrections suppressed by inverse powers of
$g_{YM}$. The corrections are actually crucial, for at any finite
value of $g_{YM}$ they imply that the theory is {\it not only} SYM. At
infinite coupling the theory {\it is} only SYM, but in the IR limit,
which is precisely what one wants.

With Lorentzian 3-algebras the result is different. Giving a vev to
the scalar $X_+^I$ reduces the theory (in its ghost-free form) to {\it
  only}\, SYM without corrections, as first noticed in
Ref.~\cite{Ho:2008ei}. In hindsight, this is a negative indication for
the usefulness of the theory. In this note we have shown that one can
go directly from SYM to the Lorentzian 3-algebra theory, clearly
demonstrating that the two are equivalent theories.

Last Monday, a new candidate theory for $N$ M2-branes was announced
\cite{Aharony:2008ug}. This theory, based on $\U(N)\times \U(N)$
Chern-Simons theory with bi-fundamental matter, appears to have the
property that, as with the Bagger-Lambert $\mathcal A_4$ theory, the Higgs
mechanism of Ref.~\cite{Mukhi:2008ux} gives a non-trivial reduction to
SYM with corrections that are suppressed by inverse powers of
$g_{YM}$. This is a positive feature. Indeed we suspect it could be
used as a proof that the theory is a correct description of multiple
M2's.

As a final point, let us observe that the Lorentzian 3-algebra
theory, while not (yet) incorporating the flow to the infrared fixed
point of SYM, may yet do so if an imaginative treatment of it is
found. Quite simply one needs to expand the theory about the vev
$\langle X_+^I\rangle =\infty$. It is not ruled out that an astute
field redefinition or other modification might make this possible.

\acknowledgments{We are grateful to James Bedford and Micha Berkooz 
for bringing the work in
Ref.~\cite{Nicolai:2003bp,deWit:2003ja,deWit:2004yr} to our attention,
and to Rajesh Gopakumar and Shiraz Minwalla for helpful discussions
and encouragement. We also thank the participants of the Monsoon
Workshop on String Theory at TIFR for their comments, and the people
of India for generously supporting our research.}

\bibliographystyle{JHEP}
\bibliography{d2d2}

\end{document}